\documentclass[prb,preprint,showpacs]{revtex4}
\usepackage{graphicx}

\newcommand{\nin}{\noindent} % no indent

\newcommand{\lth}{\ensuremath{\it{l}}}
\newcommand{\mstag}{\ensuremath{m^{+}}}
\newcommand{\jaf}{\ensuremath{J_{AF}}}
\newcommand{\jint}{\ensuremath{J_{int}}}
\newcommand{\zc}{\it{z}}

\begin{document}
\title{Exchange enhanced anisotropy in ferromagnetic/antiferromagnetic
  multilayers: dynamic consequences }%of exchange enhanced anisotropy.}

\author{R. L. Stamps}
\email{stamps@physics.uwa.edu.au}
\affiliation{
School of Physics, University of Western Australia,
35 Stirling Highway, Crawley WA 6009, Australia
}
\author{K. D. Usadel}
\email{usadel@lx12.thp.uni-duisburg.de}
\affiliation{
Theoretische Physik, Universit{\"a}t, Duisburg-Essen, 47048 Duisburg, Germany
}

\date{\today}
\begin{abstract}
The phenomena of exchange anisotropy is well known in terms of static
magnetization properties such as enhanced coercivity and magnetization
loop shifts. These effects are primarily associated with effective
anisotropies introduced into the ferromagnet by exchange coupling with
a strongly anisotropic antiferromagnet. These effective anisotropies
can be understood as manifestations of a more fundamental exchange
induced susceptibility. We show that a consequence of this view is
that a class of unusual dynamic effects associated with the exchange
susceptibility should also exist. The effects become apparent near the
ordering temperature of the antiferromagnet and affect domain wall
velocities, domain wall resonances, and precessional switching of the
ferromagnet.
\end{abstract}

%75.70.-i Magnetic properties of thin films, surfaces, and interfaces
\maketitle

Exchange anisotropy is a term coined to describe the enhancement of
magnetic anisotropies in a ferromagnet through contact with an
anisotropic antiferromagnet.\cite{meiklejohn:1956} Suitably field cooled, it is possible to
prepare a ferromagnet/antiferromagnet structure in such a way as to
observe a variety of static and quasi-static magnetic properties
associated with exchange anisotropy.\cite{nogues:1999,berkowitz:1999} One of the most well known
phenomena is the shift of magnetization loops called exchange
bias. Enhanced coercivity is also observed, and can be distinct from
the bias shift.\cite{sumryev:2003} Both the bias and coercivity require a strong contact
exchange interaction between the ferromagnet and antiferromagnet.

The common way to understand exchange anisotropy is as an effective
anisotropy originating in the antiferromagnet. An antiferromagnet with
strong anisotropy will affect the magnetic properties of an adjacent
ferromagnet if the interface spins of the two materials are correlated
through exchange interactions. This gives rise to an effective
magnetic anisotropy in the ferromagnet that can have unusual symmetry
properties upon reversal of the ferromagnet magnetization. Typically
these properties can be understood by identifying reversible and
irreversible magnetization processes in the
antiferromagnet.\cite{kouvel:1960,malozemoff:1988,stiles:1999,suess:2003}
%In this
%paper we are primarily concerned with effective anisotropies
%associated with the reversible processes, and show how these processes
%can lead to a class of interesting dynamic phenomena in the
%ferromagnet.

%In this letter we argue that this concept of an effective magnetic
%anisotropy in the ferromagnet due to its exchange coupling to the
%antiferromagnet is also valid for a number of interesting dynamic
%problems providing a clear separation of time and length scales is
%possible. To be more specific we argue that the degrees of freedom of
In this letter we argue that this concept of an effective magnetic
anisotropy in the ferromagnet due to its exchange coupling to the
antiferromagnet is also valid for a number of interesting dynamic
problems for which there is a clear separation of time and length
scales. To be more specific we argue that the degrees of freedom of
the antiferromagnet can be integrated out leading to an effective
Hamiltonian for the ferromagnet modified by interface energies which
can be expressed by a susceptibility tensor. These are quite general
results which are a generalisation of our previous work \cite{scholten:2005}
to dynamical problems. The conditions under
which such an approach is valid is a separation of time and length
scales in the sense that the antiferromagnet is in (local) thermal
equilibrium responding to the slow dynamics of the ferromagnet. Such a
situation is met in a number of important problems, ranging
from basic research to applications. Examples are domain wall
dynamics, ferromagnetic resonance or switching of single domain
ferromagnetic particles, to name just a few.

%It was recently shown that the effective anisotropies due to
%reversible processes in the antiferromagnet can be understood on a
%more fundamental level in terms of exchange induced
%susceptibilities.
The key idea is to realize that the ferromagnet is
only affected by exchange coupling across the interface to a magnetic
moment somehow induced at the interface of the antiferromagnet. The
magnitude of the moment is determined by the exchange coupling across
the interface and any applied external magnetic fields.\cite{nowak:2002}  The
corresponding effective field acting on the ferromagnet can be defined
locally at each lattice site. At site $i$ along the interface on an
atomistic level it is given by
\begin{equation}
h_{\alpha}(i) = -\frac{\partial}{\partial S_{\alpha}(i)}
\mathcal{H}_{o} + {\jint}\sigma_{\alpha}(i),
\label{eq:h_eff}
\end{equation}
\nin where $\mathbf{\sigma}$ is the nearest neighbour spin across the
interface on the antiferromagnet side, $\mathcal{H}_{o}$ is the
hamiltonian representing all other energies affecting spins within the
ferromagnet including the external applied field and magnetic
anisotropies and $\alpha$ denotes cartesian components. %For a ferromagnet with a Curie temperature large (as
%compared to the Neel temperature of the antiferromagnet), thermal
%fluctuations of the ferromagnet are neglegible. 
This effective field determines the dynamics of the ferromagnet. If
this dynamics is slow enough and if the fields acting on the
antiferromagnet vary slowly in space the antiferromagnet stays in (local)
thermal equilibrium so that a thermal average restricted to the
antiferromagnet can be performed in Eq.(\ref{eq:h_eff}) resulting in
an effective field acting on the ferromagnet given by 
%\begin{equation}
%\frac{1}{2}\left( h_{\alpha}(i+\delta)+h_{\alpha}(i) \right)\approx-\frac{\partial}{\partial 
%S_{\alpha}(\rr)}
%\mathcal{H}_{o} + \jint m_{\alpha} =h_{\alpha}(\rr).
%\label{eq:heff}
%\end{equation}
\begin{equation}
\tilde{h}_{\alpha}(i) =-\frac{\partial}{\partial 
S_{\alpha}(i)}
\mathcal{H}_{o} + \jint m_{\alpha}(i) 
\label{eq:heff}
\end{equation}
\nin where $m_{\alpha}(i)$ is the thermally averaged interface
magnetization on the antiferromagnet (AFM) side. This effective field
can be used for instance in the Landau-Lifshitz-Gilbert equations for
the spins in the ferromagnet determining their dynamics under the
assumption of slow dynamics as stated above. %The key assumption
%leading to this equation is a separation of time and length scales as
%explained above which has to be checked in applications.

To proceed we assume linear response to be valid in which case the
reversible part of the antiferromagnet interface magnetization 
is given by
\begin{equation}
\it{m}_{\alpha}(i) =
J_{int}\chi_{\alpha\beta}^{(1)}S_{\beta}(i)+\mu_{o}\chi_{\alpha\beta}^{(2)}H_{\beta}
\label{eq:mm}
\end{equation}
where summation over double appearing indices is understood.
The first term describes the response to the interlayer 
exchange coupling $J_{int}$, and the second term is the response to an
external applied magnetic field $\mathbf{H}$. A third term of the form
$\mathbf{m}_{irr}(i)$ has to be added to Eq. (\ref{eq:mm}) for a disordered
antiferromagnet representing contributions from irreversible magnetic
moments pinned at the interface. This will be discussed later.

The susceptibilities $\chi_{\alpha\beta}^{(1)}$ and
$\chi_{\alpha\beta}^{(2)}$ in general differ because the exchange
interaction only affects the AFM interface layer while the external
field is applied to all AFM layers. %when %disorder is present in the
%antiferromagnet, 
These susceptibilities represent the response of the AFM interface
layer to the effective fields.
 
%The effective field due to the coupling is an interface effect
%which scales with the number of ferromagnetic layers, $\lth$.
%Suppose a fully compensated interface such that
%at a
%neighbouring site $i+\delta$ along the interface of the ferromagnet couples to another sublattice of the antiferromagnet.
%Assuming that all energies vary slowly from one lattice site to the next, one can go to a continuum representation and
%write the
%effective field acting at point $\rr$ as

%\begin{equation}
%\frac{1}{2}\left( h_{\alpha}(i+\delta)+h_{\alpha}(i) \right)\approx-\frac{\partial}{\partial 
%S_{\alpha}(\rr)}
%\mathcal{H}_{o} + \jint m_{\alpha} =h_{\alpha}(\rr).
%\label{eq:heff}
%\end{equation}

The linear response as written in Eq. (\ref{eq:mm}) is a general form
valid for situations in which the local effective fields vary slowly
on atomic length scales and for which the time scales in the
FM layer and the AFM layer are separated in the
sense that processes in the FM layer are slow as compared to those
in the AFM layer, which itself remains locally in thermal
equilibrium.  An important insight can be gained by examining the
effective free energy of the FM layer corresponding to
Eq. (\ref{eq:heff}) in which the AFM degrees of freedom are integrated
out
\begin{eqnarray}
F &  = &  \mathcal{H}_{o} 
-\sum_{i} \left(   \jint
S_{\alpha}(i)\chi^{(1)}_{\alpha\beta}\right)\mu_{o}H_{\beta} \label{eq:ff} \\
& & \nonumber
-\frac{1}{2}\jint^{2}\sum_{i} \left( S_{\alpha}(i)\chi^{(2)}_{\alpha\beta}S_{\beta}(i)
\right) .
%+S_{\alpha}(i) m_{irr,\alpha}
\end{eqnarray}
\nin This result shows how the exchange coupling of a FM layer to an
AFM layer can be taken care of quite generally by introducing
appropriate interface energy terms to the FM layer. This effective free
energy for the ferromagnet can be used as starting point for
static as well as dynamical properties under the conditions specified above.

An important consequence follows immediately from Eq. (4): the
coupling to the AFM layer introduces in the AFM interface layer a
shift in the applied field proportional to % We will show
%below how the 
$\chi^{(1)}_{\alpha\beta}$ %term can enhance the response
%to an applied field, 
and a term proportinal to $\chi^{(2)}_{\alpha\beta}$ resulting in an
enhancement of the uniaxial anisotropy. These enhancements are
proportional to the interlayer coupling $\jint$, and arise from the
reversible part of the induced magnetization at the
ferromagnet/antiferromagnet interface. Most importantly, the
enhancements can change essential symmetries associated with rotations
of the ferromagnetic magnetization. In particular, the term quadratic
in $S_{\alpha}(i)$ results in an induced uniaxial anisotropy in the FM
interface layer providing $\chi^{(2)}_{\alpha\beta}$ is not
isotropic. The maximum lowering of symmetry occurs when only one
component of the susceptibility tensors is nonzero. This can arise
when the antiferromagnet has such strong anisotropy that it behaves
like an array of Ising spins. An example are Co/CoO multilayers
studied in connexion with exchange bias.

This case of a two
sublattice Ising antiferromagnet exchange coupled to a ferromagnet
provides a useful example that affords a straightforward
analysis. Suppose that the two antiferromagnetic sublattices for an
AFM monolayer considered for simplicity lie along the $x$
direction. Further suppose that there is no disorder so that the
system has translational invariance. The thermal averaged sublattice
magnetizations aligned colinear with the $x$ axis are defined as
$a(i)$ and $b(i)$ located at position $i$
along the interface.  Each thermal average depends on the local field
acting on the sublattice. The mean field expression for $a$ is
\begin{equation}
a = \tanh \left[ \beta\left(  \jaf \zc(\mstag - m) + H + \jint
  S_{x}(i) \right)\right]
\label{eq:mf}
\end{equation}
\nin where $\beta$ is the inverse temperature, $\zc$ is the
coordination number for the monolayer, $S_{x}(i)$ is the x component of
the interface ferromagnet component, $m=(a+b)/2$ is the induced
magnetization in the $x$ direction, and $\mstag=(a-b)/2$ is the
staggered magnetization. A similar expression can be found for the $b$
sublattice. If the induced magnetization is small compared to the
staggered magnetization, the susceptibility defined by $m=\chi
(H+\jint S_{i})$ is
\begin{equation}
\chi = \frac{\beta}{\beta \jaf z + \cosh^{2}(\beta \jaf \zc \mstag)}.
\end{equation}
Its maximun is at the Neel temperature $T_N=J_{AF}z$ and given by
%\begin{equation}
$\chi_{max} = \frac{1}{2T_N }$.
%\end{equation}
\nin The staggered magnetization is determined by the zero field,
 uncoupled average $\mstag=\tanh(\beta \jaf \zc \mstag)$.  Note that
 without disorder $\chi$ is independent of position, but that $m$ can
 be a function of position through $S_{x}$. The temperature dependence
 of $\chi$  and $\mstag$ are shown in Fig. 1 for
a square lattice monolayer with $\zc=4$.  The maximum in the
susceptibility occurs when $\mstag$ vanishes. Above the critical
temperature the system is paramagnetic and the induced magnetization
remains strongly susceptible.

Now suppose that the free energy without
exchange coupling to the antiferromagnet is of the form 
%\begin{equation}
%\mathcal{H}_{o}(\rr)=-\mu_{o}HS_{x}(\rr)-DS_{x}^{2}(\rr)+\mathcal{H}_{ex}(\mathbf{S}(\rr)).
%\label{eq:ho}
%\end{equation}
 \begin{equation}
 \mathcal{H}_{o}  =  \sum_{j} \left(
 -\mu_{o}HS_{x}(j)-DS_{x}^{2}(j) \right)+\mathcal{H}_{ex}.
\label{eq:ho}
\end{equation}
 \nin Here, a uniaxial anisotropy with strength $D$ and axis along the
$x$ direction is assumed. An external field is aligned along the easy
direction and an exchange energy indicated by $\mathcal{H}_{ex}$ is
included for completness.  When the contribution corresponding to the
exchange induced effective field is included, the energy keeps the
same form but with new parameters
 \begin{equation}
\tilde{\mathcal{H}}_{o}  =  \sum_{j} \left(
 -\tilde{\mu}_{o}HS_{x}(j)-\tilde{D}S_{x}^{2}(j) \right)+\mathcal{H}_{ex}
\label{eq:hotilde}
\end{equation}
providing we neglect variations of $S(i)$ across the FM
layer. Otherwise Eq.(\ref{eq:ho}) has to be supplemented by the
interface terms appearing in Eq.(\ref{eq:ff}).  
 
The new parameters appearing in the Hamiltonian,
Eq. (\ref{eq:hotilde}), are an enhanced permeability
\begin{equation}
\tilde{\mu}_{o}=\mu_{o}[1+(\jint/\lth)\chi]
\end{equation}
\nin and an enhanced anisotropy 
\begin{equation}
\tilde{D}=D+[\jint^{2}/(2\lth)]\chi,
\end{equation}
\nin both strongly
dependent on temperature through $\chi$. These parameters are averaged
over the film thickness and incorporate any surface effects. Here $l$
denotes the number of FM layers.
The magnitude of the enhancements of the parameters $\tilde{\mu}_{o}$
and $\tilde{D}$ can be estimated in terms of the coercive field
$h_{c}$ defined by Stoner Wohlfarth switching. One can show
that\cite{scholten:2005}
\begin{equation}
\jint\chi/\lth=\frac{h_{c}/\jint-2D/\jint}{1-h_{c}/\jint} .
\end{equation}
\nin A reasonable estimate for the interface exchange is the value of the antiferromagnetic exchange. The
corresponding exchange field is much larger than $h_{c}$. If $h_{c}/\jint\ll 1$, then the anisotropy enhancement
obeys $\tilde{D}/D\approx h_{c}/(2D)$.

 Example values of $D$ and $\jaf$ corresponding roughly to typical
materials were used in Ref. (\cite{scholten:2005}) in order to
calculate $h_{c}$ (in energy units). These numbers are for a
relatively weak anisotropy in the ferromagnet, so that the maximum
value of $h_{c}/\jint$ is on the order of $0.1$ for $2D/\jint=0.02$.
This corresponds to $(\tilde{\mu_{o}}/\mu_{o})_{max}\approx 1.1$ and
$(\tilde{D}/D)_{max}\approx 5$.

These maxima occur exactly at the Neel temperature of the
antiferromagnet in the case of an antiferromagnetic monolayer, or very
near in temperature if more antiferromagnetic layers are present. At
this critical temperature the coupled field is highly susceptible and
strong enhancements affecting phenomena that depend upon $\tilde{D}$
are expected. 

For a disordered antiferromagnet the situation is more
complicated. The problem is that for zero field cooling, the
antiferromagnet can be in equilibrium or in a frozen state. In the
latter case, field cooling can result in exchange bias effects. In the
former case, the AFM exchange interaction is obtained by replacing
$J_{AF}$ by $J_{AF}x$ where $x$ denotes the concentration of magnetic
sites in the antiferromagnet. This is valid within an effective
medium approach where $T_N(x)=J_{AF}x$ is the Neel temperature
obtained within this approximation. The maximum of the susceptibility
is at $T_N(x)$ and is given by
%\begin{equation}
$\chi_{max} = \frac{x}{2T_N(x) }$
%\end{equation}
showing that the concentration of magnetic sites cancels. Thus we
 arrive at the important result that the maximal susceptibility is
 unaffected by dilution (within an effective medium approach) while
 the temperature at which this appears decreases linearly with
 $x$.\cite{scholten:2005} In the case of an exchange bias upon field
 cooling, an irreversible antiferromagnet interface magnetization has
 to be added in Eq. (3). The corresponding bias field then appears as
 a contribution to the external field.

%The linear response formalism allows one to see clearly associated
%effects on dynamical magnetic properties. Provided that the dynamics
%of the ferromagnet occur over time scales much different from those of
%thermal fluctuations taking place in the antiferromagnet as discussed
%above,. One can see from $\chi$ directly the
%enhancement at the critical temperature as shown in Fig. 1.

An important application of our results concerns the dynamics of
domain walls in a ferromagnet coupled to an antiferromagnet. In this
case there are at least three quantities that govern phenomena for
which the separation of time scales is valid: the domain wall width
$\Delta$, domain wall energy $E_{DW}$, and domain wall mass
$m_{DW}$. Using well known expressions for each of
these,\cite{aharoni:2001} the above arguments can be used to estimate
the maximum enhancements of each quantity in terms of the maximum
coercivity. In particular one finds that the wall width decreases with
the enhancement according to
$(\tilde{\Delta})_{max}\approx\sqrt{2A/(h_{c})_{max}}$ while the wall
energy and wall mass increase as
$(\tilde{E}_{DW})_{max}\approx\sqrt{A(h_{c})_{max}}$ and
$(\tilde{m}_{DW})_{max}\sim\sqrt{(h_{c})_{max}/(2A)}$. In all of these
expressions the exchange energy of the ferromagnet is $A$.

One can immediately appreciate the effect of these enhancements by
considering the effects on domain wall velocity.  A Bloch or Neel wall
in a ferromagnetic film without defects will move at a finite velocity
when an external magnetic field is applied. The velocity $v_{DW}$ is
proportional to the product of the driving field $\mu_{o}H$ and the
domain wall width. The constant of proportionality is determined by
gyromagnetic and damping parameters.

Now suppose that the ferromagnet is exchange coupled to an
antiferromagnet. The wall velocity is typically small enough to
satisfy the separation of time scales needed to apply the above
enhancement theory. The modified wall velocity is then proportional to
the modified permeability and wall width according to
$\tilde{v}_{DW}\sim\tilde{\mu}_{o}\tilde{\Delta}$. Although the
coupling to the field is enhanced by the exchange coupling, with $D/J_{int}\ll 1$ the
largest effect is the reduction of wall width. This means that the
wall velocity slows dramatically at the critical temperature. The
effect is analogous to the increased effective mass of an electron due
to the creation of a cloud of virtual phonons while moving through a
crystal. Here the domain wall is a moving inhomogeneity in the local
field acting on the antiferromagnet, and thereby generates a response
which is largest at the critical temperature where the susceptibility
$\chi$ peaks. The result is an increase in the effective 'inertia' of
the ferromagnetic domain wall and corresponds directly to the
enhancement of the domain wall mass.

It is important to recognize that the modifications to the domain
wall velocity discussed so far depend only upon enhanced coercivity,
and not exchange bias. Additional effects due to exchange bias will
appear as an overall increase or decrease of the velocity that is
independent of applied field magnitude. Exchange bias in this theory
is due to irreversible regions of the antiferromagnet coupled to the
ferromagnet. The contribution to the effective field is, as noted
above, of the form $J_{int}m_{irr}$ and will thereby provide an
additional effective field driving the domain wall.% The sign of the
%field is determined by the sign of $J_{int}$.

Other dynamic phenomena in the ferromagnet are likewise affected. For
example, the wall energy affects thermal nucleation rates in the
ferromagnet. The wall energy determines the effective energy barrier
in an Arrhenius law for the rate,\cite{braun:1990} so that the energy
enhancement due to coupling to the ferromagnet will provide an
exponential decrease to the nucleation rate. In other words, if the
nucleation rate is $n$, then $\ln{n}/\beta\sim-\tilde{E}_{DW}$ and
there will be dramatic slowing of the nucleation rate near the
critical temperature.

As another example, a pinned ferromagnetic domain wall will exhibit a
resonance due to restoring torques associated with the pinning
centre.\cite{slonc:1979} As in the case of simple harmonic motion, the
resonance frequency depends on the ratio of restoring force to
effective mass. The frequency is usually at least one order of
magnitude smaller than the ferromagnetic resonance frequency, so the
separation of time scales requirement is fulfilled for the exchange
coupled ferromagnet and antiferromagnet. The enhancement of the domain
wall mass in the ferromagnet will cause a lowering of the pinned wall
resonance frequency.

Finally, the enhancement should also be apparent in the field induced
switching of single domain ferromagnetic particles exchange coupled to
an antiferromagnet. The precession limited switching rate of a
Stoner-Wohlfarth particle, for example, depends critically upon the
anisotropy barrier.\cite{bauer:2000} The situation is somewhat
complicated when different possible orientations of the driving field
are considered, but enhancements of the effective anisotropy barrier
will be the dominate factor determining minimum switching fields and
the required pulsed field duration.
 
In summary, induced exchange anisotropy in a ferromagnet due to
exchange coupling to an antiferromagnetic layer have been argued to
have a number of inportant dynamical effects besides modifying
coercivity. Slow dynamical processes that occur on time scales longer
than thermal fluctuation times near the Neel temperature can be
dramatically modified by exchange induced anisotropies. These
processes include domain wall motion, resonance, and magnetic reversal
and precession limited switching. The dynamical enhancements arise
from processes that are associated primarily with reversible
magnetization in the antiferromagnet, and are not related directly to
exchange bias shifts.

\begin{acknowledgments}
We thank the Australian Research Council and the Deutsche
Forschungsgemeinschaft (SFB 491) for %partial
support of this work.
\end{acknowledgments}

\newpage
% %%%%%%%%%% chi figure
\begin{figure}
\label{fig:chi}
\includegraphics[width=8cm]{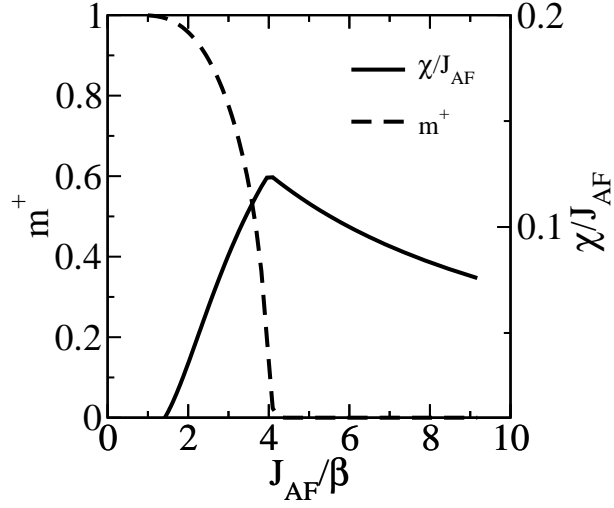}
\caption{Susceptibility $\chi$ of induced magnetization at the interface with an antiferromagnetic monolayer with
$\zc=4$.
Also plotted is the staggered magnetization $\mstag$. The susceptibility has a maximum when the staggered
magnetization vanishes.}
\end{figure}
% %%%%%%%%%%
\end{document}